\documentclass[12pt]{article}
\usepackage{amsmath,amssymb,bbm,amsthm}
\usepackage{graphicx}
\usepackage{subcaption}
\usepackage{url} 
\usepackage{xcolor}
\usepackage[toc,page]{appendix}
\usepackage{float} 

\usepackage{multicol,multirow}

\usepackage{hyperref}

\usepackage{natbib}  
\newcommand{\M}{\mathcal{M}}

\newcommand{\iid}{\overset{iid}{\sim}}
\newcommand{\ind}{\overset{ind}{\sim}}

\newcommand{\mb}[1]{\boldsymbol{#1}}

\title{Spatio-Temporal Disaggregation with Changing Areal Boundaries}
\author{
Noah Ripstein\textsuperscript{1},
Patrick Brown\textsuperscript{1}
and Jamie Stafford\textsuperscript{1}\\
\textsuperscript{1}Department of Statistical Sciences, University of Toronto
}
\date{}
\begin{document}

\maketitle

\begin{abstract}
    Small area estimation and disease mapping increasingly rely on areal data where reporting boundaries change over time. We develop a computationally efficient spatio-temporal disaggregation method to recover high-resolution risk surfaces from observed counts under changing boundaries. Our approach extends the spatially aggregated log-Gaussian Cox process and uses the Extended Latent Gaussian Model framework for fast approximate posterior inference. We replace standard lognormal polygon-specific effects with gamma-distributed overdispersion which yields a marginal negative binomial likelihood, and removes one latent variable per polygon-time pair. We illustrate the approach by mapping mortality risk across shifting NUTS-3 boundaries in Belgium and the Netherlands. For the purpose of dissemination we use Codex to leverage the methodology presented in this paper for the analysis of a separate data set concerning the city of Manchester. The methodology is implemented in the open-source \textsf{R} package \textsc{DAST}. 
\end{abstract}

\section{Introduction}

In many epidemiological studies incidence of disease is reported over a common geographic region, or map, at distinct time intervals determined by, for example, a periodic census. Incidence counts are often reported within predefined observational regions, such as counties or public health administrative areas, where the boundaries of these regions can change over time. An example of this includes county borders in the United States that are redrawn every 10 years. We refer to this as the multiple map problem, where spatial disaggregation, the process of estimating the underlying intensity at a higher resolution than the available data, is often of interest. Spatial disaggregation has been applied in disease mapping \citep{LiYe2012LGCp, utazi2019spatial, brown_root-gaussian_2021}, ecology \citep{Keil2013, MurphyKilianJ.2023Badr}, economics \citep{vidoli2022health, FujiiTomoki2024Sdop}, and numerous other fields \citep{fendrich2022scalable}. 

The Besag-York-Molli\'e (BYM) model \citep{BesagJulian1991Birw} is a widely used framework for areal data, particularly popular in epidemiology and public health for smoothing localized disease rates. ~\citet{LiYe2012LGCp} showed through simulation that a spatially-aggregated log Gaussian Cox process (LGCP) consistently outperforms BYM, however, they are often fit using bespoke Markov Chain Monte Carlo (MCMC) algorithms \citep{LiYe2012LGCp, taylor_bayesian_2015, johnson2019spatially}, which are slow to run. Integrated Nested Laplace Approximations \citep{RueHåvard2009ABif} offer a faster alternative to MCMC for a class of latent Gaussian models but aggregated LGCPs have a more complex dependence structure and do not fall into the LGM class. To address this, \citet{StringerAlex2023FSAt} introduce a class of Extended Latent Gaussian Models (ELGMs) which allow complex dependence structures including spatially-aggregated LGCPs. Simultaneously, \citet{NandiAnitaK.2023d:AR} implemented a spatially aggregated LGCP in the \textsc{Disaggregation} R package. Both reduced computation from days with MCMC to minutes. However, neither considered the use of spatially aggregated LGCPs over time where data are reported through multiple maps of the same geographic region. This is the focus of the current manuscript.

In \S 2 we develop the modelling framework for an aggregated LGCP suited to multiple maps where we use a single, shared latent field across maps with varying areal boundaries. We give some attention to the treatment of overdisperion in incidence counts. Rather than attaching a Gaussian effect to each observation region to account for overdispersion (as in \citet{NandiAnitaK.2023d:AR} and \citet{StringerAlex2023FSAt}), we model extra-Poisson variation via gamma mixing. This formulation yields a negative binomial likelihood and preserves the LGCP-based disaggregation structure while avoiding map-indexed latent variables. 

Computational aspects for the implementation of the model are considered in \S3. Here sparsity is sacrificed for dimension reduction to ease computation where gamma mixing plays a role as does the use of basis functions to model the common latent field. In \S 4 we present a case study that concens mapping early childhood mortality in Belgium and the Netherlands. For the purpose of dissemination, we demonstrate in \S 5 how modern AI tools can allow the reader to use the methods of this paper in the analysis of a separate data set concerning the city of Manchester. Finally, \S 6 concludes with limitations, extensions, and practical recommendations.

\section{An Extended Latent Gaussian Model} \label{section:methods}
Denoting a spatial region of interest by $\M$, consider a sequence of inhomogeneous Poisson processes $\{\mb{P}_i,~i\in [I]\}$ over time. Event locations are unknown and we instead observe counts $Y_{ij}=\mb{P}_i(S_{ij})$ for disjoint observation regions $S_{ij}$ where $\cup_{j} ~S_{ij} = \mathcal{M},~\forall i$ with $j\in [J_i]$. The regions $S_{ij}$ are typically predefined as counties, or by postal codes, where the boundaries may change over time. This setting is similar to that of \citet{StringerAlex2023FSAt}, however, we allow for an arbitrary number of processes rather than just one. This affects both the methodological and computational developments in what follows.

Consider an aggregated log-Gaussian Cox process for the observed counts where we have
\begin{eqnarray*}
Y_{ij}|\lambda_i(\cdot),\nu_i(\cdot)&\ind&\text{Poisson}\left[ \nu_{i}(\mb{s})\int_{S_{ij}}{\cal O}_i(\mb{s})\lambda_i(\mb{s})d\mb{s}\right], \\
\log\lambda_i(\mb{s}) &=& \mb{x}_i(\mb{s})^T \mb{\beta} +\mb{u}(\mb{s}) \\
\mb{u}(\mb{s}) &\sim & \mathcal{GP}\left\{0,{M}_{\upsilon}(\cdot;\sigma,\rho)\right\} \\
\text{Cov}\{\mb{u}(\mb{s}+\mb{h}), \mb{u}(\mb{s})\} &=& \text{M}_{\upsilon}(||\mb{h}||; \sigma, \rho), \mb{s}\in\M,\mb{s}+\mb{h}\in\M\\
\nu_{i}(\mb{s})=\nu_{ij},~\mb{s}\in S_{ij}&~\text{with}~& \nu_{ij}\iid f_\tau(\cdot)
\end{eqnarray*}
Here ${\cal O}_i(\boldsymbol{s}) \ge 0$ is a known population offset, $\mb{x}_i(\mb{s})$ are observed covariates with random effects $\mb{\beta} \sim {\cal N}\left(\boldsymbol{0},\boldsymbol{\Sigma}_{\mb{\beta}}\right)$, and both ${\cal O}_i(\boldsymbol{s})$, $\mb{x}_i(\mb{s})$ can change over time. We assume a common latent spatial process $\mb{u}(\mb{s})$ and model it as a Gaussian Process with a Matérn covariance function $M_{\upsilon}(\cdot;\sigma,\rho)$ with $\upsilon$ known. Finally, $\nu_{ij}$ are region specific random effects that address potential overdispersion in the counts. 
\subsection{Discretization of the model}
To render the model computationally tractable we consider a discretization of each $\lambda_i(\cdot)$ over a finer grid of cells $ \mathcal{Q}_\ell$, each with center $ {q}_\ell$, where $\cup_{\ell}^L\,\mathcal{Q}_\ell = \mathcal{M}$ and $\mathcal{Q}_\ell \cap \mathcal{Q}_t=\emptyset$ for $\ell\neq t$. We then approximate the risk surface over this grid by $\lambda_i(\mb{s})\approx\lambda_i(q_\ell)$ when $\mb{s}\in {\cal Q}_\ell$. This induces similar approximations for $\mb{x}_i(\mb{s})\approx \mb{x}_i(q_\ell)$ and $\mb{u}(\mb{s})\approx\mb{u}(q_\ell)$, respectively, again when $\mb{s}\in {\cal Q}_\ell$. Note covariates may be known at a much finer resolution than the observed counts, critical to disaggregation \citep{NandiAnitaK.2023d:AR}, and the above discretization accommodates that. Defining ${\boldsymbol U}=\{\mb{u}(q_\ell),\ell\in [L]\}$, an index set ${\cal L}_{ij}=\{\ell: S_{ij}\cap {\cal Q}_\ell\neq\emptyset\}$ that captures the intersection of an observation region with all grid cells, and expected counts as ${\cal O}_{ij\ell} = \int_{\mathcal{Q}_\ell \cap S_{ij}} {\cal O}_i(\boldsymbol{s})d\boldsymbol{s}$ the above model becomes
\begin{eqnarray*}
Y_{ij}|\lambda_{{\cal L}_{ij}},\nu_{ij}&\ind&\text{Poisson}\left[ \nu_{ij}\sum_{\ell\in {\cal L}_{ij}} {\cal O}_{ij\ell}\lambda_i(q_\ell)\right], \\
\log\lambda_i(q_\ell) &=& \mb{x}_i(q_\ell)^T \mb{\beta} +\mb{u}(q_\ell) \\
\boldsymbol{U}|\theta&\sim&  {\cal N}\left[\boldsymbol{0},\boldsymbol{\Sigma_{\sigma\rho}}\right]\\
\nu_{ij}&\iid&f_\tau(\cdot)
\end{eqnarray*}
where the entries of $\boldsymbol{\Sigma}_{\sigma\rho}$ are determined by the Matérn covariance function above. As noted in \citet{StringerAlex2023FSAt} it is now evident from the index set ${\cal L}_{ij}$ why the above model belongs to the ELGM class given $|{\cal L}_{ij}|>1$ reflecting a more complex dependence structure than the simpler LGM class where $|{\cal L}_{ij}|=1$.

\subsection{Treatment of overdispersion}
In what follows, it's convenient to denote the collection of latent variables, or parameters, in the model as $\mb{\omega}=\{\mb{\beta},\mb{U},\mb{\nu}\}$ where $\mb{\nu}=\{\nu_{ij}, i\in [I],j\in[J_i]\}$. Here we note that the dimension of $\mb{\omega}$ depends on the sample size due to the presence of overdispersion for each observation region $S_{ij}$. A key insight of \citet{StringerAlex2023FSAt} was noting this leads to both theoretical and computational concerns. For aggregated log-Gaussian Cox processes, it is common to model overdispersion with a log-normal distribution - an approach popularized by the highly influential BYM model of \citet{BesagJulian1991Birw} and adopted widely. (See, for example, \citet{liu2018using, michal2021bayesian} or \citet{NandiAnitaK.2023d:AR}). This also includes \citet{StringerAlex2023FSAt} surprisingly given the dimension of $\mb{\omega}$ remains a concern in that work.

To address this, we alternatively model overdispersion with a gamma distribution, ${\nu_{ij}}\sim \text{Gamma}(\tau^{-2},\tau^{-2})$ where for either formulation of $f_\tau(\cdot)$ we have $\text{Var}(\nu_{ij})= \tau~\forall i,j$. The resulting Poisson-gamma mixture can be simplified via analytical integration so the observed counts follow a negative binomial distribution. In this reformulation the latent variable becomes $\mb{\omega}=\{\mb{\beta},\mb{U}\}$ with reduced dimension as we have removed the need to compute a latent overdispersion variable $\boldsymbol \nu$. The likelihood above can now be conveniently replaced with
\begin{equation}
\label{eq:negbin-Y}
Y_{ij} \mid \lambda_i(\cdot), \tau \overset{\text{ind}}{\sim} \text{NegBin} \left(\sum_{\ell\in {\cal L}_{ij}} {\cal O}_{ij\ell}\lambda_i(q_\ell), \tau^2 \right).
\end{equation}

\subsection{A Basis Function Representation}
The dimension of $\mb{\omega}$ can be reduced further if we consider a basis expansion for the common latent surface $\mb{u}$. Following a finite element representation advocated by \citet{lindgren_explicit_2011}, we approximate $\mb{u}$ as
$$
\mb{u}({q}_\ell)
\approx \sum_{m=1}^d \phi_m({q}_\ell)\,{\gamma}_m 
$$
where \(\{\phi_j(\cdot)\}\) are known, piecewise-linear basis functions defined on a triangular mesh that covers \(\mathcal M\). Commonly referred to as tent functions due to their shape, each basis function, $\phi_j$ is defined to be 1 at the $j^{th}$ vertex, 0 at every other vertex, and linear interpolation among adjacent vertices determines the value of the function otherwise. By design, the mesh weights $\boldsymbol{\gamma}=\{\gamma_m,m\in[d]\}$ then give the value of the approximation for the set of vertices in the mesh and have a Gaussian distribution with variance matrix 
 $\boldsymbol{\Sigma}_\gamma$ determined by the Matérn covariance function.
 We now have $\mb{\omega}=\{\mb{\beta},\mb{\gamma}\}$ and dimension reduction results from $\text{dim}(\mb{\gamma})=d<L$. Further details about this approach can be found in \citet{lindgren_explicit_2011}.

\subsection{Specification of the prior}

We adopt the Penalized Complexity (PC) framework \citep{simpson_penalising_2017} to assign principled priors to the model hyperparameters, denoted collectively as $\boldsymbol\theta = \{\sigma, \rho, \tau\}$. For example, for the spatial Matérn field, we use the PC priors developed by \citet{fuglstad_constructing_2019}, which penalize deviation from a flat, constant base model by favoring small values of the marginal standard deviation $\sigma$ and large values of the spatial range $\rho$.

For the extra-Poisson variation, we are motivated by the PC prior framework in constructing a prior that shrinks the negative binomial model toward a base Poisson model. As shown in the Appendix, this results in an exponential prior on the dispersion parameter $\tau$. This is analogous to the exponential prior on the Gaussian random effect used by \citet{NandiAnitaK.2023d:AR} and \citet{StringerAlex2023FSAt} in their model.

By specifying a flatter prior for $\tau$ relative to $\sigma$, the model is encouraged to attribute residual variance to independent overdispersion before introducing spatial complexity. In what follows, we continue to use the notation $\boldsymbol\theta=\{\sigma,\rho,\tau\}$ for the collection of covariance parameters.

\section{Approximate Bayesian inference} \label{sec:approx-inf}
The above model specifications identify the joint distribution $\pi(\mb{\omega},\mb{y},\mb{\theta})$ where $\mb{y}$ simply denotes the collection of all observed counts. For inference purposes the posterior $\pi(\mb{\omega}|\mb{y})$ is of interest and, within the ELGM class, its approximation involves several elements. These include a Gaussian approximation for the posterior distribution $\pi(\mb{\omega}|\mb{y},\mb{\theta})$
$$
\widetilde\pi_G(\boldsymbol \omega \mid \boldsymbol y,\boldsymbol\theta)
\equiv \mathcal N\!\big(\hat{\boldsymbol \omega}_{\boldsymbol\theta},\,\boldsymbol H_{\hat{\boldsymbol w}_{\boldsymbol\theta}}(\boldsymbol\theta)^{-1}\big),$$
where  $\hat{\boldsymbol \omega}_{\boldsymbol\theta}$ is the mode of $\pi(\mb{\omega},\mb{\mb{y}},\mb{\theta})$ and $\boldsymbol H_{\hat{\boldsymbol \omega}_{\boldsymbol\theta}}(\boldsymbol\theta)$ the Hessian, and a Laplace approximation $\widetilde\pi_L({\boldsymbol\theta}|\boldsymbol y)$ for the posterior of $\boldsymbol \theta$. Numerical integration over $\mb{\theta}$ gives an approximation for $\pi(\mb{\omega}|\mb{y})$
$$
\widetilde\pi(\boldsymbol \omega\mid \boldsymbol y)=\sum_{z\in{\cal Q}}
\widetilde\pi_G(\boldsymbol \omega\mid \boldsymbol y,\widetilde{\boldsymbol\theta}_z)~\widetilde\pi_L(\widetilde{\boldsymbol\theta}_z|\boldsymbol y)~\omega _z
$$
where quadrature points $z$ and weights $w_z$ are determined by an adaptive Gauss-Hermite rule known for sound theoretic properties. The approximation $\widetilde\pi_L({\boldsymbol\theta}|\boldsymbol y)$ itself depends on the same quadrature points and weights. Full details are given in 
\citet{StringerAlex2023FSAt} and we compute these approximations using several components:
\begin{itemize}
\item Automatic Differentiation(AD): Template Model Builder is used for coding the joint density function
$\pi(\mb{\omega},\mb{\mb{y}},\mb{\theta})$ including various design matrices. The use of automatic differentiation (AD) tracks derivatives of the individual components of the log density and produces first and second derivatives using the chain rule. This, for example, determines the Hessian $\boldsymbol H_{\boldsymbol \omega}(\boldsymbol\theta)$.
\item Inner Optimization:  Using the gradients and Hessians computed by AD, an ``inner" optimization step is carried out with Trust Region optimization \citep{nocedal2006numerical} to produce the mode $\hat{\boldsymbol \omega}_{\boldsymbol\theta}$. Evaluating the Hessian at ths mode then gives the Gaussian approximation $\widetilde\pi_G(\boldsymbol \omega\mid \boldsymbol y,\boldsymbol\theta)$.  
\item Outer Optimization: Approximating the joint distribution $\pi(\mb{\mb{y}},\mb{\theta})$ as 
$$
\widetilde{\pi}(\mb{\mb{y}},\mb{\theta})={\pi(\hat{\mb{\omega}}_\theta,\mb{\mb{y}},\mb{\theta})\over\widetilde\pi_G(\hat{\mb{\omega}}_\theta\mid \boldsymbol y,\boldsymbol\theta)}
$$
and again using derivatives produced by AD and Trust Region optimization, an ``outer" optimization step determines the mode of $\widetilde{\pi}(\mb{\mb{y}},\mb{\theta})$, denoted as $\hat{\mb{\theta}}$.
\item Adaptive Gaussian Hermite Quadrature: A set of representative points around the mode $\hat{\mb{\theta}}$, namely $\{\widetilde{\mb{\theta}}_z, z\in {\cal Q}\}$, and weights $\{w_z,z\in{\cal Q}\}$ are determined by adaptive Gaussian Hermite Quadrature. These are then used to to compute the Laplace approximation $\widetilde\pi_L(\widetilde{\boldsymbol\theta}_z|\boldsymbol y)$ as
$$
\widetilde\pi_L(\boldsymbol\theta|\boldsymbol y)={{\widetilde{\pi}(\mb{\mb{y}},\mb{\theta})}\over{\sum_{z\in{\cal Q}}
\widetilde\pi(\boldsymbol y,\widetilde{\boldsymbol\theta}_z)~\omega_z}}
$$
and finally $\widetilde\pi(\boldsymbol \omega\mid \boldsymbol y)$.
\end{itemize}
Our implementation builds on the excellent \texttt{disaggregation} R package \citep{NandiAnitaK.2023d:AR}. We use its mesh construction and SPDE infrastructure (via \texttt{fmesher}) without modification. To support multiple maps across time, we extend the functions for constructing design matrices to handle a common latent field with time-varying areal boundaries. We likewise adapt their \texttt{TMB} template \citep{kristensen2016tmb} to aggregate over multiple maps and use the marginal negative binomial likelihood. Unlike the \texttt{disaggregation} package which fits the model specified by the \texttt{TMB} template through Empirical Bayes with a single Laplace approximation, we fit using the ELGM fitting algorithm via the \texttt{AGHQ} package \citep{aghq_pkg_stringer}. The ELGM algorithm with a single quadrature point reduces to a Laplace approximation, and increasing the number of quadrature points yields a fully Bayesian algorithm which is more accurate than the Emperical Bayes approach when compared with MCMC \citep{StringerAlex2023FSAt}. A full set of code and reproducible examples are contained in the GitHub repository at {\tt github.com/nripstein/DAST}. An open-source R package, \textsc{DAST} (DisAggregation in Space and Time), provides a user-friendly interface similar to the \textsc{Disaggregation} package.

\section{Mortality in Belgium and Netherlands}

\begin{figure}[htpb]
\begin{center}
\begin{subfigure}{0.32\linewidth}
\includegraphics[width=\textwidth]{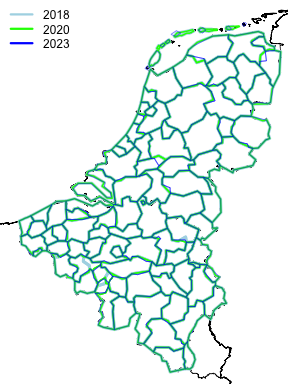}
    \caption{Boundaries}
\end{subfigure}
\begin{subfigure}{0.32\linewidth}
\includegraphics[width=\textwidth]{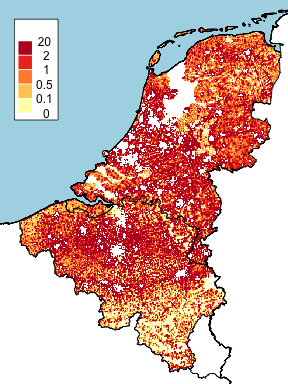}
    \caption{Population 2018}
\end{subfigure}
\begin{subfigure}{0.32\linewidth}
\includegraphics[width=\textwidth]{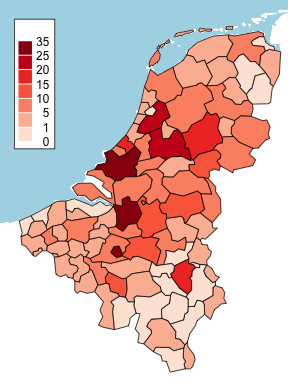}
    \caption{Deaths 2018}
\end{subfigure}
\begin{subfigure}{0.32\linewidth}
\includegraphics[width=\textwidth]{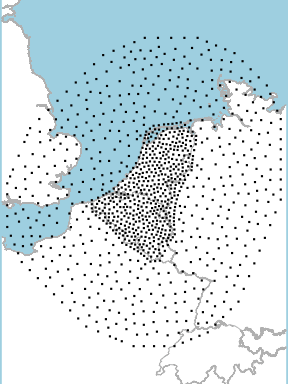}
    \caption{Basis function vertices}
\end{subfigure}
\begin{subfigure}{0.32\linewidth}
\includegraphics[width=\textwidth]{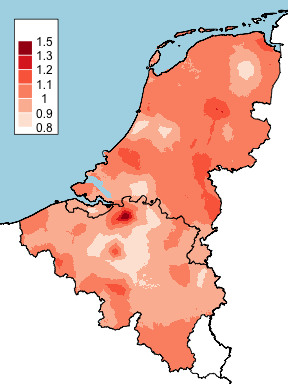}
    \caption{Relative risk}
\end{subfigure}
\begin{subfigure}{0.32\linewidth}
\includegraphics[width=\textwidth]{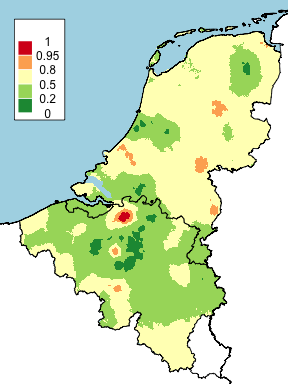}
    \caption{Exceedance}
\end{subfigure}

\caption{Boundaries of NUTS 3 regions, posterior median of relative risk 
$\exp[U(s)]$ 
and posterior exceedance probabilities $\text{pr}[U(s) > 0 | Y]$.
}
\label{fig:deathResults}
\end{center}
\end{figure}

\begin{figure}[htpb]
\begin{center}
\begin{subfigure}{0.32\linewidth}
\includegraphics[width=\textwidth,trim = 0cm 0cm 6mm 0cm]{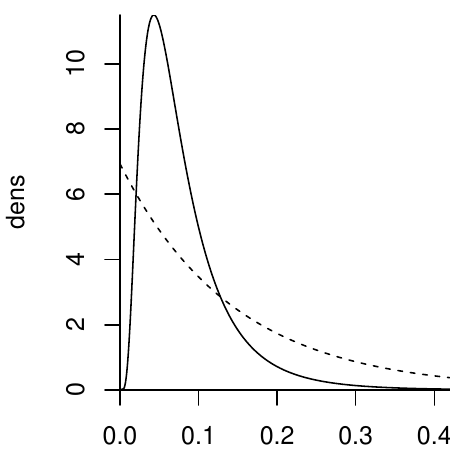}    \caption{Observation SD $\tau$}
\end{subfigure}
\begin{subfigure}{0.32\linewidth}
\includegraphics[width=\textwidth,trim = 3mm 0cm 3mm 0cm]{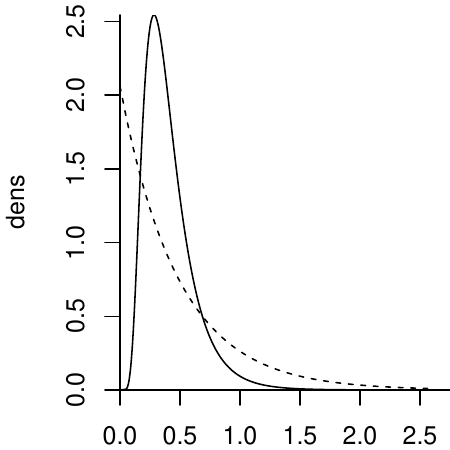}
    \caption{Spatial SD $\sigma$}
\end{subfigure}
\begin{subfigure}{0.32\linewidth}
\includegraphics[width=\textwidth,trim = 6mm 0cm 0mm 0cm]{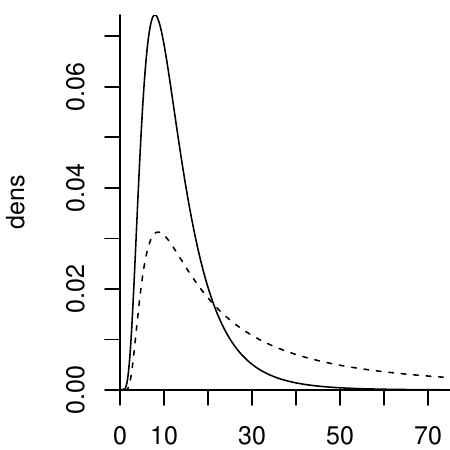}
    \caption{Range (km) $\phi/1000$}
\end{subfigure}
\caption{Prior distributions ( - - - ) and posteriors ( \rule[0.5ex]{2em}{0.8pt} ) for model parameters.}
\label{fig:deathParam}
\end{center}
\end{figure}

To demonstrate the practical utility and computational efficiency of the DAST framework, we apply the model to a high-resolution mortality study across international borders. 

Figure \ref{fig:deathResults} shows the results from applying the methodology to data on deaths in Belgium and the Netherlands from 2018 to 2023.  The data modelled are death counts for Females under the age of 5 at the NUTS 3 level, from the Eurostat dataset \verb!demo_r_magec3!. Yearly population data on a 1km grid are from the Worldpop product R2025A.  A full set of code, including specific web URL's for data sources, are available in the Github repository {\tt github.com/nripstein/DAST-ppr-code}.

The NUTS 3 administrative boundaries underwent revisions in both 2020 and 2023. By using the shared latent field $u(\mathbf{s})$ within our framework, we are able to integrate data across these three distinct map configurations into a single, spatially-consistent posterior risk surface. Figure \ref{fig:deathResults}b shows the population density in 2018 along with vertices of the spatial basis functions
and Figure \ref{fig:deathResults}c shows the observed number of deaths in 2018.   
Figure \ref{fig:deathResults}d shows the vertices of the basis functions used to approximate the Gaussian random field.

The Negative Binomial model was fit using 2 quadrature points and exponential priors on the overdispersion standard deviation $\tau$, the spatial standard deviation $\sigma$ and the scale parameter $1/\phi$.  
Parameter priors and posteriors are shown in Figure \ref{fig:deathParam}.  
These exponential priors encourages small values, with prior modes at zero, which encourage a smooth spatial surface with low variance and strong spatial dependence.  The posterior median of relative risk appears in  Figure \ref{fig:deathResults}e. The posterior exceedance probabilities, $\text{pr}(\lambda(s) > 1 | Y)$
is in Figure \ref{fig:deathResults}f.  There is a high risk region at the border between the countries, near the Belgian city of Antwerp, where risk is predicted to be 30\% above average and there is 95\% confidence that risk is above what is typical.

\section{AI-Assisted Practitioner Workflow}

To demonstrate the accessibility of the \textsc{DAST} package for applied researchers, we present an analysis scaffolded via modern artificial intelligence tools. This example highlights a workflow enabling practitioners straightforwardly adopt our methodology for novel, real-world analyses when dealing with shifting administrative boundaries.

\begin{figure}[ht]
\subfloat[Log relative risk\label{fig:pooled-maps-1}]{\includegraphics[width=0.49\linewidth]{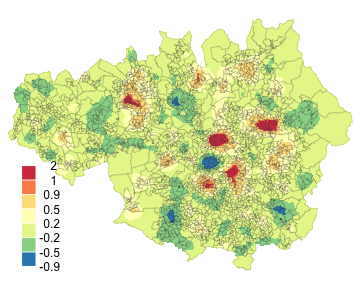} }\subfloat[Exceedance probability\label{fig:pooled-maps-2}]{\includegraphics[width=0.49\linewidth]{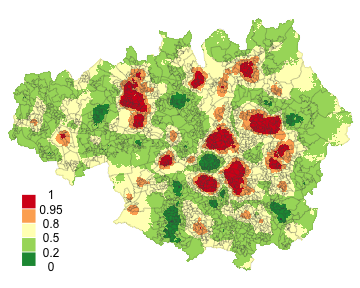} }\caption[Result maps for Manchester ]{Result maps for Manchester : posterior median of log relative risk E$[\mb{u}(\mb{s}) | \mb{y}]$ and exceedance probability pr$[\mb{u}(\mb{s}) | \mb{y}]$.}\label{fig:man-maps}
\end{figure}

Figure \ref{fig:man-maps} shows the log relative risk and exceedance probabilities for under-five mortality in Manchester, UK, from 2008 to 2012. Death counts are available for Lower Layer Super Output Areas (LSOAs), the boundaries of which changed in 2010. Code for the analysis was generated by the Codex AI tool from the following prompt, which references the \texttt{disagg.Rmd} file used for the Belgium and Netherlands example:

\begin{quote}
I would like to replicate the analysis in the file `disagg.Rmd` to estimate the risk of under-5 mortality in Manchester, UK for 2008 to 2012 using death counts from the ONS at the LSOA level and population from worldpop.  I know the 2008 and 2009 data are on one set of boundaries and 2010 onwards are on a different set.  Please write an RMD file to download and assemble the data in the correct form.  Also include code similar to disagg.Rmd to run the model.
\end{quote}

While the resulting code required a moderate amount of debugging and modification, it successfully established the data pipeline and model fitting procedure. The final source document producing an extended set of results is available on this paper's GitHub repository ({\tt github.com/nripstein/DAST-ppr-code})

\section{Discussion} \label{section:discussion}
We have introduced a scalable spatio-temporal disaggregation framework for areal data where regional boundaries change over time. 
By replacing Gaussian polygon-specific effects with gamma-distributed overdispersion, we marginalize the Poisson-gamma hierarchy to a negative binomial likelihood, eliminating one latent variable per polygon-timepoint pair. 
This structural change keeps the dimension of the inner Hessian in approximate posterior inference fixed with respect to the number of polygons and enables efficient fitting for datasets with thousands of polygon-time pairs.

Relative to existing approaches, our method extends the work of \citet{StringerAlex2023FSAt} and \citet{NandiAnitaK.2023d:AR} by combining three capabilities: 
(i) the accuracy of LGCP-based disaggregation, 
(ii) the ability to handle changing boundaries across time, and 
(iii) the dimensionality reduction achieved by modeling extra-Poisson variation though gamma-mixing, and the computational efficiency of our approximate inference approach.

There are several limitations to note. 
First, we assume that polygon-level overdispersion is adequately modeled by gamma mixing; in settings with heavier-tailed or asymmetric extra-Poisson variation, alternative mixing distributions may be preferable. 
Second, the current formulation shares a single latent spatial field $u(\mathbf{s})$ across time, without explicit temporal dynamics; this may be restrictive when spatial correlation structures evolve over time. 
Third, as with all SPDE-based methods, performance depends on mesh resolution and discretization: coarse meshes reduce computational cost but may undersmooth fine-scale features, while fine meshes increase accuracy at the expense of computation.

Several extensions follow naturally. 
Incorporating temporal correlation in $u(\mathbf{s}, t)$, such as separable or nonseparable spatio-temporal Mat\'ern models, could further improve efficiency in multi-timepoint settings. 
Allowing time-varying spatial range or variance parameters would accommodate evolving spatial dependence. 
Extensions to other likelihoods, such as zero-inflated models would broaden applicability to data with excess zeros.

By aligning the model structure with the strengths of modern deterministic approximation methods, our approach narrows the gap between fine-scale spatial inference and computational feasibility in complex, real-world applications. 
We hope that the accompanying open-source \textsf{R} package \textsc{DAST} will facilitate adoption of these methods in applied domains where large, irregular, and evolving areal datasets are the norm.\\

\noindent\textbf{Acknowledgments:}
We are grateful to our colleague, Rohan Alexander, for encouraging us to consider the use of modern AI tools for the dissemination of these methods. We also acknowledge the support of CANSSI Ontario and the Natural Sciences and Engineering Research Council of Canada (NSERC).

\clearpage
\section*{Appendix: Negative Binomial Penalized Complexity prior}\label{ap:PC-prior}

We construct a penalized complexity (PC) prior that shrinks a negative binomial model toward a Poisson base model. We parameterize
\[
Y \mid \mu,\alpha \sim \mathrm{NegBin}(\mu,\alpha),\qquad
\mathbb{E}[Y]=\mu,\ \ \mathrm{Var}(Y)=\mu+\alpha\,\mu^2,\quad \alpha=\tau^2.
\]
The Poisson base model corresponds to $\tau=0$.

We represent the negative binomial model as a Poisson–Gamma mixture:
\[
Y \mid v \sim \mathrm{Poisson}(v\,\mu),\qquad
v \sim \mathrm{gamma}\!\left(\tau^{-2},\,\tau^{-2}\right),
\]
so that $\mathbb{E}[v]=1$ and $\mathrm{sd}(v)=\tau$. The Poisson base model corresponds to $\tau=0$, in which case $v \equiv 1$.

Let $(Y,v)$ denote the data $Y$ and the mixing variable $v$. 
Under our model with overdispersion parameter $\tau$, the joint density is
\[
p_\tau(y,v) \;=\; \mathrm{Poisson}(y \mid v\,\mu)\;\mathrm{gamma}\!\left(v \,\middle|\, \tau^{-2},\, \tau^{-2}\right),
\]
and under the Poisson base model ($\tau=0$) we have
\[
p_b(y,v) \;=\; \mathrm{Poisson}(y \mid \mu)\;\mathrm{gamma}\!\left(v \,\middle|\, \tfrac{1}{b^{2}},\, \tfrac{1}{b^{2}}\right),
\]
where $b\downarrow 0$ makes the gamma distribution concentrate at $v=1$. The chain rule for KL divergence gives

\begin{align*}
\mathrm{KL}\!\left(p_\tau \,\|\, p_b\right)
&= \underbrace{\mathbb{E}_{v \sim \mathrm{gamma}(\tau^{-2},\,\tau^{-2})}
\,\mathrm{KL}\!\bigl(\mathrm{Poisson}(v\mu)\,\|\,\mathrm{Poisson}(\mu)\bigr)}_{\text{(A)}} \\
&\quad+\;
\underbrace{\mathrm{KL}\!\left(\mathrm{gamma}\!\left(\tau^{-2},\,\tau^{-2}\right)\,\bigg\|\,\mathrm{gamma}\!\left(b^{-2},\,b^{-2}\right)\right)}_{\text{(B)}}.
\end{align*}

Term (A) depends on $\mu$ and reflects how the data model changes with $v$, 
while term (B) measures how the mixing distribution deviates from its base. 
Following the PC prior construction, we focus on term (B) since it measures 
the complexity of the overdispersion component independent of the mean.

We now evaluate term (B) in the limit $b \downarrow 0$, which corresponds to the base model $v \equiv 1$. The KL divergence between $\mathrm{gamma}(\tau^{-2},\tau^{-2})$ and $\mathrm{gamma}(b^{-2},b^{-2})$ is
\begin{equation} \label{eq:KL-gammas}
\begin{split}
\mathrm{KL}(\tau^{-2} \,\|\, b^{-2})
&= (\tau^{-2} - b^{-2})\,\psi(\tau^{-2})
- \log \Gamma(\tau^{-2}) + \log \Gamma(b^{-2}) \\
&\quad + b^{-2}\bigl[\log \tau^{-2} - \log b^{-2}\bigr] + b^{-2} - \tau^{-2},
\end{split}
\end{equation}
where $\psi(\cdot)$ is the digamma function and $\Gamma(\cdot)$ is the gamma function.

To take the limit $b \downarrow 0$ (i.e., $b^{-2} \to \infty$), we use Stirling’s approximation for $\log \Gamma(b^{-2})$,
\begin{equation} \label{eq:stirling-loggamma}
\log \Gamma(b^{-2}) 
= \left(b^{-2} - \tfrac12\right)\log b^{-2} - b^{-2} + \tfrac12 \log(2\pi) + \mathcal{O}\!\left(b^2\right).
\end{equation}
Substituting \eqref{eq:stirling-loggamma} into \eqref{eq:KL-gammas} and simplifying yields
\begin{equation} \label{eq:KL-limit}
\mathrm{KL}(\tau^{-2} \,\|\, b^{-2})
= b^{-2}\!\left[\log \tau^{-2} - \psi(\tau^{-2})\right] 
- \tfrac12 \log b^{-2} 
+ \mathcal{O}(1) + \mathcal{O}(b^2).
\end{equation}

The leading term grows linearly in $b^{-2}$. Thus, as $b \downarrow 0$, the leading term of the KL divergence is dominated by
\[
\mathrm{KL}_{\mathrm{dom}}(\tau) \;=\; C\,\big[\log(\tau^{-2}) - \psi(\tau^{-2})\big],
\]
up to a constant factor $C$ from the $b^{-2}$ term in \eqref{eq:KL-limit}.

Following \citet{simpson_penalising_2017}, the constant is dropped, and the PC prior distance measure is defined as
\begin{equation} \label{eq:PC-distance}
d(\tau) \;=\; \sqrt{ 2\,\big[\log(\tau^{-2}) - \psi(\tau^{-2})\big] }.
\end{equation}
For small $\tau$, using the first-order expansion of the digamma function,
$\psi(\tau^{-2}) = \log(\tau^{-2}) - \tfrac{\tau^{2}}{2} + \mathcal{O}(\tau^{4})$,
we obtain $d(\tau) = \tau + \mathcal{O}(\tau^{3})$, so $d(\tau)$ is asymptotically equal to the standard deviation of the gamma mixing distribution. We therefore use

\[
d(\tau) = \tau
\]

Placing an exponential prior with rate $\lambda$ on $d(\tau)$ gives
\[
\pi(\tau) = \lambda \exp(-\lambda \tau).
\]
This exponential prior on the standard deviation of the gamma mixing distribution is analogous to the exponential prior on the standard deviation for Gaussian random effects \citep{simpson_penalising_2017}.

The hyperparameter $\lambda$ can be selected via a user-defined probability statement on the tail of the prior. We specify an upper bound $U$ for the overdispersion standard deviation $\tau$ and a small tail probability $p$ such that $\mathrm{Pr}(\tau > U) = p$. Given the exponential prior on $\tau$, this yields$$\lambda = -\frac{\log(p)}{U}.$$This translates directly to the variance parameterization $\alpha = \tau^2$, allowing a practitioner to intuitively constrain the prior based on the maximum plausible amount of overdispersion expected in the data apriori.
\clearpage

\bibliography{references}

\end{document}